# Superconductivity in Room Temperature Stable electride and high-pressure phases of alkali metals

Hideo Hosono[a,b,*] Sung-Wng Kim,[c,*] Satoru Matsuishi,[a] Shigeki Tanaka,[d] Atsushi Miyake,[d] Tomoko Kagayama[d] and Katsuya Shimizu[d]

[a] Frontier Research Center, Tokyo Institute of Technology, 4259 Nagatsuta, Midori, Yokohama 226-8503, JAPAN

[b] ACCEL Program, Japan Science and Technology Agency (JST), Hon-cho, Kawaguchi 332-0012, JAPAN

[c] Department of Energy Science, Sungkyunkwan Unversity, Suwan 440-746, Korea

[d] Center for Quantum Science and Technology under Extreme Conditions , Osaka University,Toyonaka 560-8531, Japan ,Japam

*Corresponding author hosono@msl.titech.ac.jp and kimsungwng@skku.edu

**ABSTRACT**

S-band metals such as alkali and alkaline earth metals do not undergo a superconducting transition (SCT) at an ambient pressure, but their high-pressure phases do. In contrast, room temperature stable electride $[Ca_{24}Al_{28}O_{64}]^{4+} \cdot 4e^-$ electride (C12A7:$e^-$) in which anionic electrons in the crystallographic sub-nanometer-size cages have high s-character exhibits SCT at 0.2-0.4K at an ambient pressure. In this paper we report that crystal and electronic structure of C12A7:$e^-$ are close to those of the high pressure superconducting phase of alkali and alkaline earth metals and the SCT of both materials is induced when electron nature at Fermi energy ($E_F$) switches from s- to sd-hybridized state.

**INTRODUCTION**

The exploration of a new high-$T_c$ superconductor and the elucidation of its mechanism are key areas of research both in material science and in condensed matter physics. Extensive efforts have yielded various superconducting materials, including cuprates[1], iron pnictides[2] and even Si[3]. In 2007, a constituent of alumina cement, heavily electron-doped $12CaO·7Al_2O_3$, referred as $[Ca_{24}Al_{28}O_{64}]^{4+}·4e^-$ (C12A7:e$^-$) electride was identified as the first superconductor based on a light metal oxide.[4] Here electride is a crystal in which electrons serve as anions[5] and may be regarded as a crystal of solvated electron. We think this discovery has two scientific meanings. One is that emergence of superconductivity arising from solvated electrons[6] was discussed from a view point of bi-polaronic mechanism [7]. The other is that this new type of superconducting materials opens up a new category of superconducting materials because light metal oxides have been regarded as typical insulators. In addition, many elements that behave as simple metals exhibit superconductivity at a high pressure accompanying with structural phase transitions[8-11]. It is noticed that various high-pressure phases of elements change to a phase with sub-nanometer-scale cavities occupied by anionic electrons[11-13]. Such a polymorth may be regarded as an elecride. Interestingly, several elemental electrides are superconductors[14,15], implying the generality of the emergence of superconducting state in metallic electrides. This view leads to a belief that elucidation of the fundamental origin responsible for the superconductivity of metallic electrides is important for finding a new superconductor.

A variety of electrical properties of C12A7 such as an insulator-conductor conversion and metal-insulator transition (MIT) originate from its three-dimensionally connected sub-nanometer-sized cage structure [16-18]. The crystal lattice of C12A7 belongs to the cubic space group $I\bar{4}3d$, and its unit cell is composed of a positively-charged lattice framework, $[Ca_{24}Al_{28}O_{64}]^{4+}$ with 12 cages and two extra-framework $O^{2-}$ ions entrapped in two out of the 12 cages as counter anions. The incorporation of electrons as counter anions instead of $O^{2-}$ ions leads to the formation of room-temperature stable electride[17] (Fig. 1**a**). Since the cages are three-dimensionally connected by sharing monolayer oxide cage walls, the energy levels for each cage interact strongly through electron tunneling to form an energy band named a "cage conduction band (CCB)," which is split from the framework conduction band (FCB) [19]. Furthermore, each empty cage has an *s*-like state, and the interaction of these states forms the CCB with an *s*-like ground state. Therefore, by the replacement of the extra-framework $O^{2-}$ ions with electrons, the CCB is partially occupied by electrons having an *s*-like nature in the cage. When a low electron concentration ($N_e$) is doped into the crystal, the *s*-electrons are localized in specific cages forming an $F^+$-like center with a localized electron. In this case, electrical conduction occurs through a thermal activation of the electron from the isolated level of an $F^+$-like center to the CCB. With an increase in $N_e$, the split level merges into the CCB to lead to switching of the conduction mechanism from hopping to band conduction, i.e. the *s*-electrons spread out over the cages [18]. Thus, the metallic C12A7:e$^-$ with itinerant *s*-electrons may be regarded as an *s*-band metal as a first approximation, with the property that the highest-energy electron is in an *s*-orbital, like for *s*-block alkali and alkaline earth metals.

It should be noted, however, that the metallic C12A7:e$^-$ undergoes the SCT at ~0.2 K at an ambient pressure. This fact is apparently incompatible with a well-known fact that an *s*-band metal of alkali and alkaline earth metals does not exhibits SCT at an ambient pressure but some of them become superconducting [20-24] through the pressure-induced change with the exception of lithium which has a very low $T_c$ (0.4 mK) at an ambient pressure [25]. This fact implies the existence of an unrevealed mechanism to explain the emergence of superconductivity in metallic C12A7:e$^-$. In this paper, we report how the sub-nanometer-sized cage network structure of C12A7:e$^-$ is responsible for the emergence of superconductivity, based on comparison with the pressure-induced superconductivity of conventional *s*-band alkali metals. An insight into superconductors is addressed with a close correspondence in crystal and electronic structures between superconductors of C12A7:e$^-$ and high-pressure phases of *s*-band alkali metals. Furthermore, a high-pressure experiment on the superconductivity of C12A7:e$^-$ verifies a similarity to superconducting high-pressure phases of *s*-band alkali metals.

**EXPERIMENTALS AND CALCULATIONS**

Superconducting C12A7:e$^-$ was prepared by using a single-crystal precursor, grown via the floating zone (FZ) technique[26]. To dope electrons into the insulating single-crystal, heat treatment under Ti metal vapor was employed. Single-crystal plates were sealed in a silica glass tube (inner volume ~ 10 cm$^3$) with Ti metal shots, followed by thermal annealing at temperatures between 800 and 1,100℃ for 12 ~ 24 hr. The extent of the replacement of free oxygen ions with electrons, i.e. electron concentration (*Ne*) was controlled by adjusting the temperature and the duration of the chemical reduction. The *Ne* values of the single crystals were estimated from optical reflectance spectra in the infrared to ultraviolet region (450 to 40000 cm$^{-1}$) measured at room temperature [28]. Electrical resistivity was measured in the temperature range 0.04 ~ 300 K by the four-probe method, using Pt electrodes deposited on the surface of Ti-treated samples. For temperatures between 2 and 300 K, a conventional $^4$He cryostat was used, while a $^3$He/$^4$He dilution refrigerator was employed in the temperature range of 0.04 ~ 5 K. The Seebeck coefficient (S) was measured in the temperature range of 2 ~ 300 K by heating each end of sample alternatively to create a temperature difference of 0.1 ~ 1 K and recording the voltage induced by this temperature gradient between the two ends. The temperatures were monitored by Chromel/Au-0.07%Fe thermocouples (Nilaco Co., Japan), attached at both ends of samples. The measured voltage was corrected by subtracting the voltage induced by the Chromel to obtain the S values of samples. To examine superconducting properties under high pressure, the AC susceptibility of single-crystal C12A7:e$^-$ was measured using piston cylinder and diamond anvil cells. The dimension of C12A7:e$^-$ used in the piston cylinder cell was 1500μm x 600μm x100μm. The pressure was determined from the measurements of the Tc change of metal Pb associated with the pressure. The primary and pickup coils were wound around both C12A7:e- and lead. The data were collected as output signals of a lock-in amplifier [29]. The hydrostatic pressure was retained with a fluid pressure transmitting medium, Daphne 7373 or Ar gas. The applied pressure in the high pressure apparatus was controlled to be constant during the measurements on cooling and heating processes, i.e. these measurements were performed always at constant pressure. The piston-cylinder and diamond anvil cells were loaded to the adiabatic demagnetization refrigerator (ADR) for the AC susceptibility measurements under high pressure.

The VIENNA ab initio simulation package (VASP)[30] was used to calculate the band structure and total and projected DOSs for metallic C12A7:e$^-$ and Li, via the projector-augmented wave method and the Pedrew-Burke-Ernzerhof form of the generalized gradient approximation functional. A plane-wave energy cutoff of 500 eV and a 2×2×2 k-mesh were used. Isosurfaces of the charge densities were drawn by using

VESTA code [31].

**RESULTS AND DISCUSSION**

First, we examine the transport properties of superconducting C12A7:e$^-$ and draw the electronic phase diagram as a function of $N_e$. Figure 1b shows the temperature dependence of normalized electrical resistivity ($\rho/\rho_{2K}$) in the range from 0.04 to 0.5 K, in which the resistivity for single-crystals (samples C and D) and thin films (samples E and F) in Ref.4 are reproduced. Each $\rho/\rho_{2K} - T$ curve exhibits an abrupt drop at a temperature of 0.09 ~ 0.40 K, indicating that the sample undergoes the SCT. The $T_c$ of the single-crystals increases from 0.09 K (sample A, 240 Scm$^{-1}$), 0.11 K (sample B, 300 Scm$^{-1}$) to 0.19 K (sample C, 770 Scm$^{-1}$) and further to 0.20 K (sample D, 810 Scm$^{-1}$) with an increase in electrical conductivity($\sigma$) at 300K. Since the $\sigma$ value of the single-crystals monotonically increases with $N_e$, we conclude that $T_c$ of C12A7:e$^-$ superconductor increases with increasing $N_e$. Figure 1c shows the electronic phase diagram as a function of $N_e$. This plot clearly demonstrates that the superconducting state appears for $N_e > 1 \times 10^{21}$ cm$^{-3}$, and this critical $N_e$ agrees with that at which the metal-insulator transition (MIT) occurs. That is, all the metallic samples exhibit superconductivity at an ambient pressure. The variation of $T_c$ with $N_e$ shows a super-linear behavior. At the same time, the Seebeck coefficient ($S$) shows a distinct sign change at this critical $N_e$.[27] Figure 1d shows $S_{300K}$ values as a function of $N_e$. As $N_e$ increases, the sign of $S_{300K}$ changes from negative to positive on going from the insulating side to the metallic side, indicating that the slope (energy derivative) of the electronic density of state (DOS) at $E_F$ is positive in the insulating side and becomes negative for the higher $N_e$. The values of $S$ in the metallic samples are positive in the whole temperature range 2 ~ 300 K.[27] These positive $S$ values provide important information about the electronic structure that the shape of DOS in the vicinity of $E_F$ is either dome-like or valley-like, as shown schematically in the inset of Fig. 1d. It is worth noting that $E_F$ of metallic C12A7:e⁻ is located at the decreasing region of DOS, and that the DOS continues to decrease with $N_e$, since all metallic samples exhibit positive $S$. According to the McMillan formula [32] based on Bardeen-Cooper-Schrieffer (BCS) theory (the SCT in C12A7:e⁻ was confirmed to be BCS-type by heat capacity measurements)[33], the Tc should decrease because SCT occurs at the metallic side. However, the observed Tc is super-linear increase with $Ne$.

Figure 2 shows variation in electronic structure with electron-doping. For understanding of this SCT at an ambient pressure and $T_c$ increase with decreasing DOS, we examined that how the electronic structure of the CCB is affected by the incorporation of electrons in the C12A7 cages. When $N_e$ is low in semiconducting C12A7, electrons form a localized band of $F^+$-like centers at ~ 0.4 eV below the bottom of the CCB[19], but when $N_e$ reaches near the $N_c$ of MIT, the localized levels start to merge with the CCB. Thus, $E_F$ is pushed up to ~ 0.9 eV above the bottom of CCB in [Ca$_{24}$Al$_{28}$O$_{64}$]$^{4+}$(4e⁻), yielding the metallic state (Fig. 2a ). **Figure 2b** describes the DOS shape near and in the CCB and the variation of $E_F$ with $N_e$. As expected from the positive $S$ values, $E_F$ in the metallic C12A7:e⁻ is located at a decreasing region of the DOS (indicated by arrow at the bottom of Fig. 2b) with a valley-like shape

The partial DOS of each conduction band (FCB and CCB) is mainly formed by the component of Ca ions and the DOS of FVB (framework valence band) is by the O ions forming the cage framework. For the detail component of CCB, as presented in Fig. 2c, the Ca 4$s$-projected DOS is dominant near the bottom of the CCB, and the contribution of Ca 3*d*-projected DOS becomes larger and dominant around the top of CCB. These overall features of the CCB remain unchanged irrespective of the incorporation of electrons in our calculations. Thus, when $E_F$ is located in the CCB, electrons in the cages change their nature from the *s*-state of localized $F^+$-like centers to the *sd*-hybridized state of Ca. These electronic features explain the exceptional SCT of metallic C12A7:e¯ at an ambient pressure, because non-*s*-electrons with the freedom of orbital multiplicity tend to form a stable Cooper pairs than *s*-electrons. Furthermore, the finding that the ratio of Ca *d*-projected DOS to Ca *s*-projected DOS increases for increasing $N_e$ in the metallic state explains the anomalous $T_c$ increase with decreasing total DOS.

Conceptually, this change of electron nature from *s*- to *sd*-hybridized state is similar to that from *s*- to a non-*s*-state under high pressure in *s*-band alkali metals, which is the main cause of superconductivity. It is thus worthwhile to compare superconductivity in metallic C12A7:e¯ with the pressure-induced superconductivity of the conventional *s*-band metals such as Li, the simplest alkali metal. It is well-known that pressure-induced superconductivity in alkali metals competes with the symmetry-breaking phase transition, e.g. Li is a highly symmetrical bcc structure at ambient pressure and room temperature and transforms to close-packed rhombohedral 9R structure[34] at low temperatures with no sign of superconductivity down to 0.4 mK. However, it undergoes sequential phase transitions with pressure from bcc to fcc, hR1, cI16, and Cmca[20-24, 35]. Among these high-pressure Li phases, the fcc, hR1, and cI16 phases exhibit the SCT with the highest $T_c$ ~ 20 K at the boundary regime between fcc and cI16 phase[20-22]. Another alkali metal, sodium also shows the pressure-induced structural transitions to cI16 phase at 103 GPa[11], and this cI16-Na phase is also expected to be superconducting near ~ 1 K[15] Surprisingly, the space group ($I\bar{4}3d$) of the high-pressure cI16-Li phase is the same as that of C12A7[14,36]. The crystal of the cI16-Li phase has a distorted bcc structure with 16 atoms per unit cell and a lattice constant of 0.5271 nm (Fig. 3**a** and 3**b**). The arrangement of Li atoms in the cI16 phase is very similar to that of the cage network structure of C12A7; the 16 Li atoms occupy a Wyckoff 16c site, which is one of the two Al-occupied sites (Wyckoff 16c and 12b) in the C12A7 case. Thus, the 8 Li atoms form a polyhedron encompassing sub-nanometer-sized free space (interatomic distance of ~ 0.34 nm), and the center of the polyhedron corresponds to a Wyckoff 12a site which is the same position as the center of the C12A7 cage. The most surprising result is that the cI16 structure of alkali metals may be regarded as an electride, i.e., a large amount of valence charge working as anions within the cage-like polyhedron rather than in close proximity to the lattice atoms.[14] That is, the naturally built-in cage network structure of C12A7:e$^-$ at an ambient pressure is a replica of the cage-like polyhedron of high-pressure superconducting cI16-Li phase.

The close structural similarity between cI16-Li and C12A7:e$^-$ is directly reflected in their electronic structures. The pressure-induced symmetry-breaking towards the cI16 phase is accompanied with the splitting of the degenerate 2*p*-orbitals and subsequently increased 2*p*-projected DOS and decreased 1*s*-projected DOS at $E_F$, inducing the change of electron nature from *s*- to *p*-state. This low-symmetry cI16-Li phase has a pronounced electron density due to the overlapping of degenerate *p*-orbitals in the crystallographic free space of the sub-nanometer-sized polyhedron generated by symmetry-breaking, as shown in Fig. 3**a** and 3**b**, whereas the electron density of ambient-pressure bcc-Li phase is mainly distributed around the lattice atoms. This peculiar density distribution of low-symmetry cI16-Li phase is also observed in metallic [Ca$_{24}$Al$_{28}$O$_{64}$]$^{4+}$(4e$^-$). Because the naturally built-in C12A7 cage structure with low symmetry provides an unoccupied *d*-state of Ca in the cage of insulating [Ca$_{24}$Al$_{28}$O$_{64}$]$^{4+}$(2O$^{2-}$) (Fig. 2c), sufficient electron doping into the cages for the MIT leads to the contribution of *d*-projected DOS at $E_F$, arising the change of electronic nature at $E_F$ from *s*- to *sd*-hybridized state. Consequently, the crystallographic free space of this cage has a large electron density, and its electron distribution is elliptical along the $S_4$ symmetry axis (Figures 3d and 3e). This result implies strongly that the Ca-sublattice, assisted by the bonding with Al and O ions in the cage network structure of C12A7, is the primary factor for the realization of electronic structure specific to the high-pressure cI16-Li phase.

Meanwhile, the superconductivity of high-pressure phase of calcium was first confirmed at 2K under 44 GPa and found to increase with elevating pressure up to 25K under 161 GPa,[24] and originated from the change of electron nature from *s*- to *d*-state at $E_F$. However, superconducting high-pressure phases of calcium show highly close-packed structures, which have a substantially shorter Ca-Ca distance [37] (from 0.38 nm at 0.1MPa to 0.33 nm at 44 GPa or to 0.22 nm at 161 GPa) than that (0.57 nm) of the Ca-sublattice in the C12A7 cag structure. This result indicates that a large electron density in the sub-nanometer-sized space within the cages and the change of electron nature from s- to non-s-state at $E_F$ are the key factors to induce the superconductivity in electrides at an ambient pressure.

Another similarity in electronic structure is found in the band structures. The calculated band structure of the cI16-Li phase agrees with that of the previous studies[38,39]. It was reported that the lowering of Fermi energy by symmetry-breaking towards the cI16-Li phase induces an electron pocket of *p*-character band associated with the Fermi surface (FS) nesting at the symmetry point *H* (Fig. 3c), where a pseudogap appears at $E_F$ (Fig.4). The overall feature of band structure of C12A7:e⁻ is similar to that of the cI16-Li phase (Fig. 3f), i.e., the valley-like DOS at $E_F$, which was proved by the Seebeck coefficient measurements, is similar to the pseudogap of cI16-Li phase. Thus, the change of electron nature from *s*- to non-*s*-state at $E_F$ and the appearance of a pseudogap probably associated with FS nesting are the fundamental characteristics of electronic structures to be correlated with the superconductivity in both electrides. Notwithstanding both have a similar electronic structures, the $T_c$ of $[Ca_{24}Al_{28}O_{64}]^{4+}(4e^-)$ is much lower than that (~ 20 K) of cI16-Li phase. This difference may be understood in terms of DOS at $E_F$ and electron-phonon coupling constant (λ). The calculated DOS($E_F$) and λ of $[Ca_{24}Al_{28}O_{64}]^{4+}(4e^-)$ are $4.5 \times 10^{21}$ eV⁻¹cm⁻³ and 0.46,[26] respectively. These values are rather lower than those of the cI16-Li phase ($18 \times 10^{21}$ eV⁻¹cm⁻³ and 0.98[38]). It is expected that the pressurization is a promising approach to raising the $T_c$ of C12A7:e⁻ by the enhancement of *sd*-hybridized state at $E_F$.

We performed AC susceptibility measurements of C12A7:e⁻ under pressure generated by using a piston-cylinder cell and diamond anvil cell up to 5.0 GPa. Figure 5 shows the pressure dependence of the $T_c$ and the temperature derivative of the critical magnetic field -d$H_{c2}$/d$T$ for C12A7:e⁻ with $N_e$ ~ $2.0 \times 10^{21}$ cm⁻³ (Sample D in Fig. 1b). $T_c$ increases with applying pressure and has a peak at around 5 GPa in consistent with the increase of d$H_{c2}$/d$T$ which is associated with the density of state. The maximum $T_c$ is 2.36 K at 5.2 GPa, about 10 times higher than that at ambient pressure. The pressure where $T_c$ starts decreasing is corresponding to the pressure where an abrupt reduction of superconducting volume fraction in ac-susceptibility is observed.

Recently, it was found that a pressure induced structural phase transition occurred at around 5 GPa by a synchrotron X-ray diffraction measurement at room temperature and high pressure[40]. The decrease of $T_c$ (and $-dH_{c2}/dT$) at high pressure can be due to the structural phase transition. The kink of the $-dH_{c2}/dT$ in #1-#3 around 3.5 GPa can be ascribed to the appearance of the high pressure phase with different pressure distributions arising from the solidification of pressure transmitting medium. This result makes a sharp contrast to the case of Li, absence of the SCT below 20 GPa until the cI16 phase is stabilized.

**SUMMARY**

We proposed an idea by noting the close similarities of the crystal and electronic structures between C12A7:e⁻ electride and high-pressure cI16-Li phase that the occurrence of SCT in C12A7:e- at an ambient pressure and anomalous $T_c$ increase with decreasing the DOS are attributed to the change of electron nature from *s*-state to *sd*-hybridized state at $E_F$. This change originates from the naturally built-in Ca-sublattice in cage network structure which encompasses sub-nanometer-sized free space. Furthermore, both C12A7:e⁻ and high-pressure cI16-Li contain large electron densities in their crystallographic free spaces. This similarity suggests us that electrides with anionic electrons with itinerant nature in the inherent sub-nanoscale free space would be a novel family of superconductors.

**References**

1. J.G.Bedorz and K.A.Muller, Z.Phys. B, 1986, **64**, 189.

2. Y.Kamihara, T.Watanabe, M.Hirano, and H.Hosono, *J. Am. Chem. Soc.* 2008, **130**,

3296.

3. E.Bustarret. E. *et al. Nature, 2006,* **444**, 465.

4. M. Miyakawa, M. et al. J. Am. Chem. Soc. 2007,129, 7270.

5. J.L.Dye, *Acc. Chem. Res.* 2009,**42**, 1564.

6. P. Edwards, *J.Supercond.*2000,**13**,933.

7. P.P.Edwards, C. N. R. Rao, N. Kumar and A. Sasha Alexandrov, *Chem.Phys.Chem.* 2006, **7**, 2015.

8. K.Shimizu *et al. Nature*, 2001, **412**, 316.

9. G.Profeta *et al. Phys. Rev. Lett.,* 2006, **96**, 047003.

10. J.J.Hamlin, V.G.Tissen and J.S. Schilling, *Phys. Rev. B.,* 2006, **73**, 094522.

11. Y.Ma *et al. Nature, 2009,* **458**, 182.

12. C.J.Pickard and R.J. Needs, *Nature Mater.,* 2010, **9**, 624.

13. M.Martinez-Canales, C.J. Pickard and R.J. Needs, *Phys. Rev. Lett.,* 2012,**108**, 045704.

14. M.Hanfland, K.Syassen, N.E.Christensen and D.L.Novikov, *Nature,* 2000, **408**, 174.

15. N.E.Christensen and D.L.Novikov, *Phys. Rev. B* .,2006,**73**, 224508.

16. K.Hayashi, S.Matsuishi, T.Kamiya, M.Hirano and H.Hosono, *Nature*, 2002, **419**, 462.

17. S.Matsuishi, *et al. Science*, 2003,**301**, 626.

18. S.W.Kim *et al*. *Nano Lett.* 2007, **7**, 1138.

19. P.V.Sushko, A.L.Shluger, K.Hayashi, M.Hirano and H.Hosono, *Phys. Rev. Lett.,2003,* **91**, 126401.

20. K.Shimizu, H.Ishikawa, D.Takao, T.Yagi, and K.Amaya, *Nature,* 2002, 4**19**, 597.

21. V.Struzhkin, I.M.Eremets, W.Gan, H.Mao and R. Hemley, *Science* 2002, **298**, 1213

22. S.Deemyad and J.S. Schilling, *Phys. Rev. Lett.* 2003, **91**, 167001..

23. N.E. Christensen and D.L.Novikov, *Phys. Rev. B*, 2006, **73**, 224508.

24. T.Yabuuchi, T.Matsuoka, Y.Nakamoto, and K.Shimizu, *J. Phys. Soc. Jpn*. 2006,75, 083703.

25. J.Tuoriniemi *et al. Nature, 2007,* **447**, 187.

26. K.Kurashige et al. *Crystal Growth & Design,* 2006, **6**, 1602.

27. S-W.Kim *et al*. *Phys. Rev. B, 2009,* **80**, 075201.

28. S.Matsuishi, S-W.Kim, T.Kamiya, M.Hirano and H.Hosono, *J. Phys. Chem. C*, 2008 **112**, 4753.

29. S.Tanaka et al. *J.Korean Phys.Soc*.2013, **63**,477.

30. G.Kresse and J.Furthmüller, *Phys. Rev. B*.,1996, **54**, 11169.

31. K.Momma and F.Izumi, *Commission on Crystallogr. Comput., IUCr Newslett.* 2006, **7**, 106.

32. W.L.McMillan, Transition Temperature of Strong-Coupled Superconductors. *Phys. Rev.* 1968, **167**, 331.

33. Y.Kohama *et al*. *Phys. Rev. B. 2008,* **77**, 092505.

34. A.W.Overhauser, *Phys. Rev. Lett.*, 1984, **53**, 64.

35. J.B. Neaton and N.W.Ashcroft, *Nature*, 1999, **400**, 141.

36. M.O'keeffe and B.G. Hyde, *Crystal structures* (Mineralogical Society of America, Washington DC, 1996).

37. H.Fujihisa, Y.Nakamoto, K.Shimizu, T.Yabuuchi and Y.Gotoh, *Phys. Rev. Lett., 2008,* **101**, 095503 .

38. Y. Yao, J. S. Tse, K. Tanaka, F. Marsiglio, and Y. Ma. *Phys. Rev. B, 2009,* **79**, 054524.

39. J.B.Neaton and N.W.Ashcroft, *Phys. Rev. Lett.,* 2001, **86**, 2830.
40. H. Hara, Y. Tomota, S. W. Kim, N. Imamura, T. Atou and H. Hosono, Special Issue of The Review of High Pressure Science and Technology, Vol. 22, p. 21 (2012), (in Japanese).

**Acknowledgements**

We thank Dr. Yoshimitsu Kohama and Dr. Masashi Miyakawa of Tokyo Institute of Technology for electrical resistivity measurements in dilution cryostat, and the thin film fabrications, respectively. This work is supported by JST Accel Program. A part of this work was supported by MEXT Element Strategy Initiative Project to form a core research center.

**Figure Legends**

**Figure 1. Crystal structure and electronic transport properties of $[Ca_{24}Al_{28}O_{64}]^{4+}(4e^-)$ electride. a**, The frame indicates a cubic unit cell $[Ca_{24}Al_{28}O_{64}]^{4+}$, composed of twelve cages. Anionic electrons are omitted for simplicity. Each cage has a free space with ~ 0.4 nm inner diameter and electron occupancy of the cage is 4/12. If a cage does not contain an anion, the wave function in the cage is composed of almost free-electron-like *s*-orbital. The expanded image showing the cage incorporating an electron and the dashed line is the $S_4$ symmetry axis. **b**, Electrical resistivities at low temperatures for metallic conducting single-crystal (samples A ~ D) and epitaxial

thin-film (samples E and F) samples. The superconducting transition is observed for all the samples. The poor S/N resolution of the single-crystal samples is due to the small current flow to suppress Joule-heating leading to a shift of $T_c$. **c**, Electronic phase diagram as a function of carrier concentration $N_e$ for single crystals (filled circles) and thin films (open circles). It is worth noting that $T_c$ increases super-linearly with $N_e$. **d**, Seebeck coefficient ($S$) at 300 K, $S_{300K}$, as a function of $N_e$ for degenerate semiconducting and metallic conducting single crystals. Semiconducting and metallic conducting regions are colored yellow and pink, respectively. The sign change in $S$ is observed at $N_e \sim 1 \times 10^{21}$ cm$^{-3}$ (above sample I), which agrees with the critical Ne value for the metal-insulator transition (MIT). The inset is the schematic illustration of DOS derived from the sign change of $S$.

**Figure 2. Electronic structure of metallic $[Ca_{24}Al_{28}O_{64}]^{4+}(4e^-)$ electride. a**, Density of states (DOS) for insulating $[Ca_{24}Al_{28}O_{64}]^{4+}(2O^{2-})$ and metallic conducting $[Ca_{24}Al_{28}O_{64}]^{4+}(4e^-)$ with $N_e = 2.3 \times 10^{21}$ cm$^{-3}$. FVB, CCB and FCB denote the framework valence band, cage conduction band and framework conduction band, respectively. $E_F$ of metallic $[Ca_{24}Al_{28}O_{64}]^{4+}(4e^-)$ is located in CCB at ~ 0.9 eV above the CCB bottom. **b**, Schematic illustration of DOS for the semiconducting and metallic C12A7:e$^-$. When $N_e$ increases to $1 \times 10^{21}$ cm$^{-3}$, the localized $s$-states of the $F^+$-like centers are merged with the CCB to lead to the MIT. In the area of metallic C12A7:e$^-$ indicated by an arrow, the DOS decreases continuously with an increase in $N_e$. **c**, Partial DOS of the CCB of insulating $[Ca_{24}Al_{28}O_{64}]^{4+}(2O^{2-})$. The Ca $s$-projected DOS is dominant around the bottom of the CCB, but Ca $d$-projected DOS becomes larger and dominant near the top of the CCB. These features are unchanged in the metallic

[Ca$_{24}$Al$_{28}$O$_{64}$]$^{4+}$(4e$^-$). Thus, the contribution of Ca $d$-projected DOS arises naturally from $N_e$ where $E_F$ is located in the CCB, i.e. the metallic state is realized. The ratio of Ca $d$-projected DOS relative to Ca $s$-projected DOS increases with the increase of $N_e$ in the region of decreasing DOS in the metallic state.

**Figure 3. Similarity in crystal and electronic structures between [Ca$_{24}$Al$_{28}$O$_{64}$]$^{4+}$(4e$^-$) electride and the high-pressure cI16-Li phase. a**, **d**, Electron density isosurfaces for electrons within the conduction band (from the bottom of the conduction band to $E_F$) for cI16-Li and metallic C12A7:e$^-$ with the composition of [Ca$_{24}$Al$_{28}$O$_{64}$]$^{4+}$(4e$^-$). Green denotes the density isosurface for electrons populated in the polyhedron of cI16-Li and the cage of C12A7:e$^-$. The color of each atom is the same as that in Fig. 1**a**. **b** and **e**. The expanded image showing the polyhedron structure in cI16-Li and the Ca-sublattice structure in the cage of C12A7:e$^-$. **c**, **f**, Band structure of cI16-Li and metallic [Ca$_{24}$Al$_{28}$O$_{64}$]$^{4+}$(4e$^-$). Compared to band structure of bcc-Li (see supplementary Fig. S3), observed are broadening of $sp$-hybridized band and lowering of the unoccupied $p$-character band below $E_F$ with a minimum at symmetry point $N$. This induces anisotropic deformation in the Fermi surface, which develops nesting at the symmetry point $H$, where a pseudogap appears at $E_F$. The overall features of the band structure of [Ca$_{24}$Al$_{28}$O$_{64}$]$^{4+}$(4e$^-$) are close to those of high-pressure cI16-Li phase with an electron pocket at symmetry point $N$.

**Figure 4**. Calculated density of states (DOS) of Li phases. a. Total DOS, b. s-projected DOS and c. p-projected DOS of ambient-pressure bcc-Li and high-pressure cI16-Li phases. The high-pressure cI16-Li phase has a higher contribution of p-projected DOS

and lower contribution of s-projected DOS at $E_F$ than those of ambient-pressure bcc-Li phase, indicating the change of electron nature from s- to p-state. Another noticeable difference is the appearance of a pseudogap (steep increase in DOS near $E_F$) at $E_F$ in cI16-Li phase, which is not present in the bcc-Li phase and other high-pressure phases. The appearance of such a pseudogap at $E_F$ in the cI16-Na phase is a characteristic of high-pressure cI16 phases of alkali metals.

**Figure 5. Superconducting properties of $[Ca_{24}Al_{28}O_{64}]^{4+}(4e^-)$ electride under high pressure.** Pressure dependence of temperature derivative of critical magnetic field $-dH_{c2}/dT$ and $T_c$ for C12A7:e$^-$ with $N_e \sim 2.0 \times 10^{21}$ cm$^{-3}$ (Sample D in Fig. **1a**) were obtained by the AC susceptibility measurements using a piston cylinder cell (1st and 2nd runs) and diamond anvil cell (3rd and 4th run).

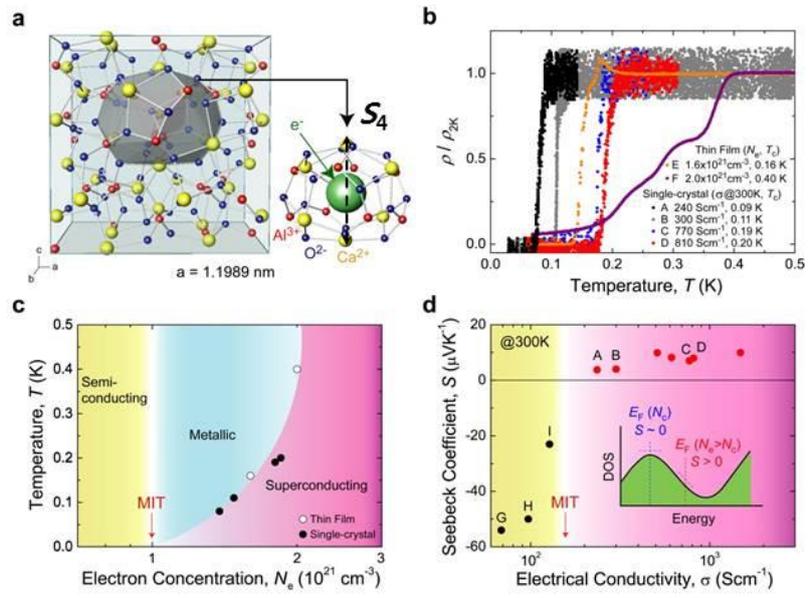

Fig 1

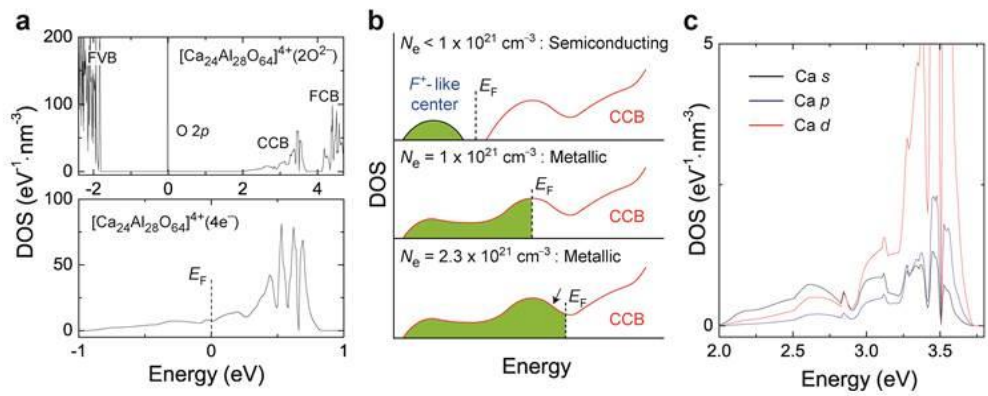

Fig 2

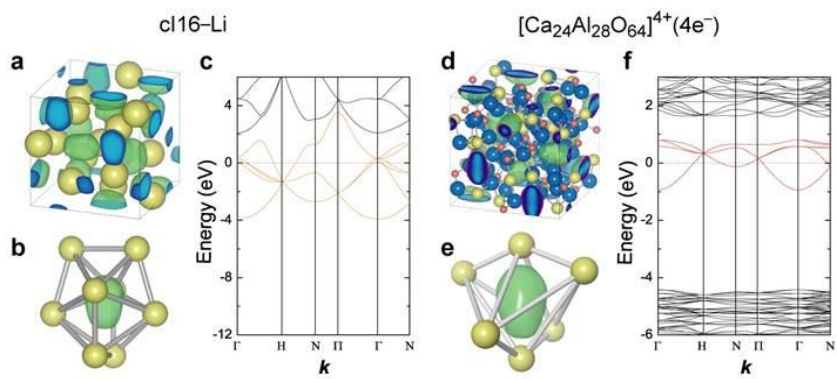

Fig 3

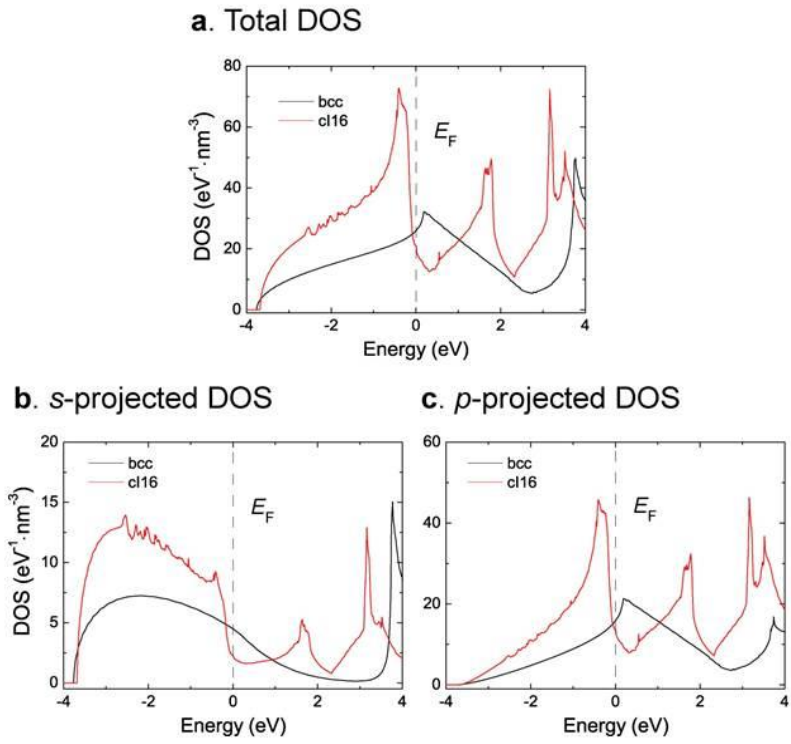

Fig 4

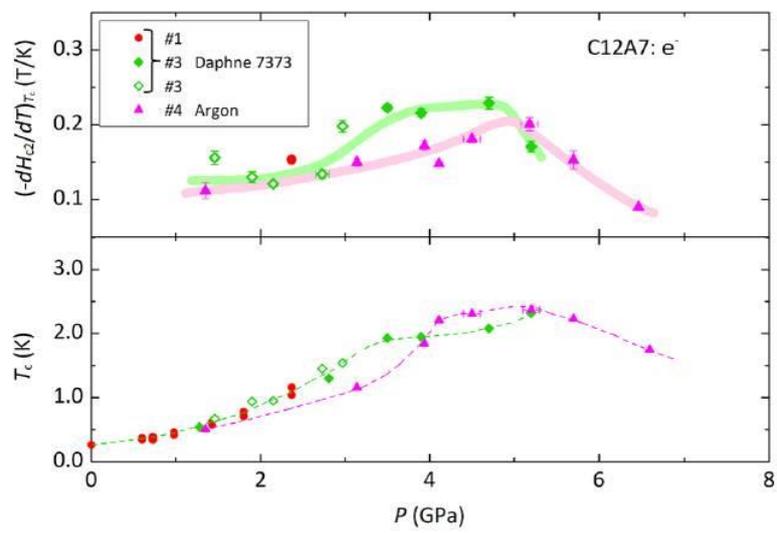

Fig 5